\documentclass{mem}
\usepackage{natbib}
\usepackage{txfonts}
\usepackage{balance}
\usepackage{graphicx}
\usepackage{flushend}
\usepackage[a4paper,breaklinks,pdftex]{hyperref}
\idline{75}{282}

\begin{document}
\def\teff{$T\rm_{eff }$}
\def\kms{$\mathrm {km s}^{-1}$}

\title{Pitfalls of AI classification of rare objects\thanks{Adapted from \citet{2022A&A...661A..52P}}}
\subtitle{Galaxy Mergers}

\author{
W.~J.~Pearson\inst{1} 
\and L.E. Suelves\inst{1}
\and S.C.-C. Ho\inst{2}
\and N.Oi\inst{3}
\and NEP Team
\and GAMA Team 
          }

\institute{
National Centre for Nuclear Research, Pasteura 7, 02-093 Warszawa, Poland \email{william.pearson@ncbj.gov.pl}
\and
Institute of Astronomy, National Tsing Hua University, 101, Section 2. Kuang-Fu Road, Hsinchu, 30013, Taiwan
\and
Faculty of Science Division II, Liberal Arts, Tokyo University of Science,	1-3, Kagurazaka, Shinjuku-ku, Tokyo 162-8601, Japan
}

\authorrunning{W.~J.~Pearson et al. }

\titlerunning{Pitfalls of AI classification}

\date{Received: Day Month Year; Accepted: Day Month Year}

\abstract{Galaxy mergers are hugely important in our current dark matter cosmology. These powerful events cause the disruption of the merging galaxies, pushing the gas, stars and dust of the galaxies resulting in changes to morphologies. This disruption can also cause more extreme events inside the galaxies: periods of extreme star formation rates and the rapid increase in active galactic nuclei activity. Hence, to better understand what goes on in these rare events, we need to be able to identify statistically large samples.
	
In the last few years, the growth of artificial intelligence techniques has seen application to identifying galaxy mergers. These techniques have shown to be highly accurate and their application has grown beyond academic studies of ``can we?'' to deeper scientific use. However, these classifications are not without their problems.
	
In this proceedings, we will explore how galaxy merger classification can be improved by adding pre-extracted galaxy morphologies alongside the traditional imaging data. This demonstrates that current neural networks are not extracting all the information from the images they are given. It will also explore how the resulting samples of rare objects could be highly contaminated. This has a knock on impact on the upcoming large scale surveys like Euclid and Rubin-LSST.

\keywords{Galaxies: interactions -- Galaxies: evolution -- Methods: data analysis -- Galaxies: statistics}
}
\maketitle{}

\section{Introduction}

Galaxy mergers underpin out current understanding of how galaxies grow and evolve. In the current dark matter paradigm, dark matter halos grow hierarchically. These halo mergers result in the galaxies in their centres also merging. This results in a single, larger dark matter halo containing a single, larger galaxy.

These interactions cause changes to the galaxy. Tidal forces involved cause material to move in the interacting galaxies. This results in morphology changes, including the creating of tidal tails or bridges. The tidal forces can also move material to the centre, triggering active galactic nuclei activity \citep[e.g][]{1985AJ.....90..708K, 2011ApJ...743....2S, 2020A&A...637A..94G} or shock the gas and trigger extreme star-formation \citep[e.g.][]{1985MNRAS.214...87J, 2013MNRAS.435.3627E, 2019A&A...631A..51P}.

Merger rates are not constant in the Universe, with more mergers found in the younger Universe. This is seen both in observations and in simulations, although there are disagreements in how the merger rate evolves \citep[e.g.][]{2007ApJS..172..320K, 2008MNRAS.386..909C, 2008ApJ...672..177L, 2011MNRAS.417.2770C, 2012ApJ...747...34B, 2013MNRAS.431.2661C, 2015MNRAS.449...49R, 2017MNRAS.464.1659Q, 2019A&A...631A..51P, 2021MNRAS.501.3215O}. These disagreements may be a result of using a fixed merger timescale when a variable timescale is more applicable \citep{2017MNRAS.468..207S}.

To better understand galaxy mergers, large samples are required. Visual identification, often considered the most reliable method, is slow and time consuming. Mergers can also be identified with morphological parameters, such as Gini or M$_{20}$ or identifying galaxies close on the sky and in redshift. With the huge volumes of data expected from upcoming surveys, such as Euclid or Rubin-LSST, machine learning has become an area of interest for merger identification.

\begin{figure}[t!]
\resizebox{\hsize}{!}{\includegraphics[]{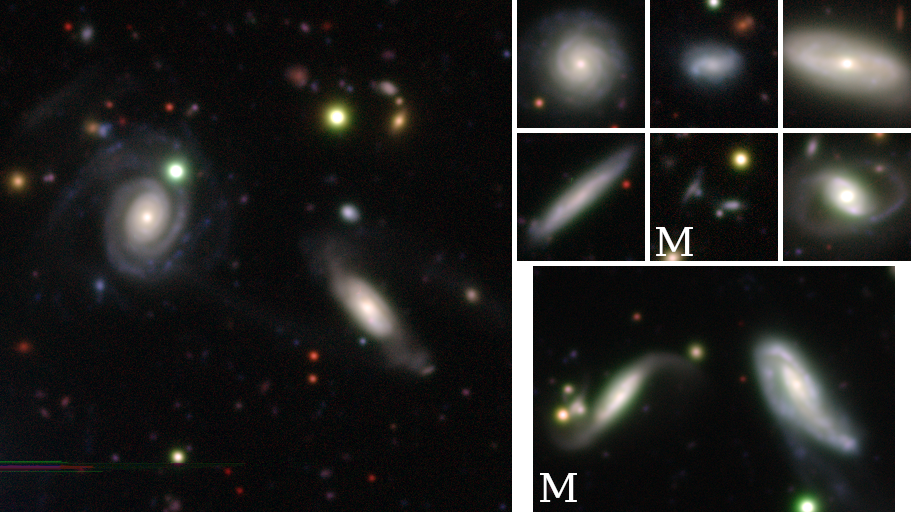}}
\caption{\footnotesize Example galaxies from HSC-NEP that will be classified as mergers or non-mergers in this work. Galaxies labelled `M' were ultimately identified as mergers in this work while the others were not.}
\label{fig:HSC}
\end{figure}

In this proceedings, we will explore how deep learning can be used to identify galaxy mergers in Hyper Suprime-Cam (HSC) Subaru Strategic Program (SSP) and North Ecliptic Pole (NEP) imaging, such as the images in Fig. \ref{fig:HSC}. It will look at how combining visual-like classification with galaxy morphologies can improve classification. It will also discuss how excellently performing neural networks can have highly contaminated samples of rare objects. As galaxy mergers are a minor class of galaxy in the universe, this misclassification problem has implications for merger studies with these upcoming surveys.

\section{Merger identification}
We aimed to identify galaxy mergers from HSC data in NEP. This was done using deep learning trained on HSC-SSP images of merging and non-merging galaxies. These were identified in the GAMA-KiDS Galaxy Zoo project and the worst miss-classifications were removed with an asymmetry-smoothness cut. This provided 1683 merging systems at $z < 0.15$ and a further 1683 non-mergers were collected.

We trained a convolutional neural network (CNN) on the r-band HSC images. This network achieved a loss of 0.473 and accuracy of 79.3\% at validation. We further trained a fully connected network (FCN) on the morphologies of the galaxies, extracted from the same r-band images used to train the CNN. The FCN achieved a loss of 0.301 and an accuracy of 88.8\% at validation. Full details of the CNN and FCN architectures can be found in \citet{2022A&A...661A..52P}.

\begin{figure*}[t!]
	\resizebox{\hsize}{!}{\includegraphics[]{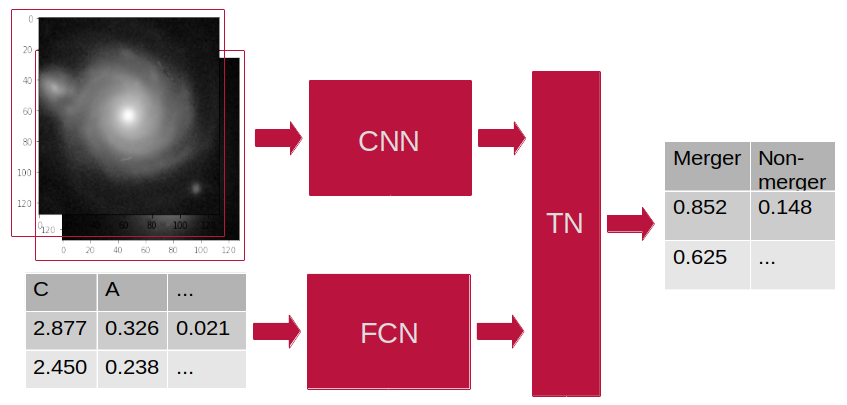}}
	\caption{\footnotesize Schematic of the combined FCN and CNN network. The CNN takes an r-band image as input while the FCN takes the 17 morphological parameters. The concatenated layer of the CNN and FCN are passed into a Top Network (TN) discriminator before a two-neurone binary classification is generated.}
	\label{fig:FULL}
\end{figure*}

As the morphologies can be considered a compression of the images, the better FCN results imply that the CNN cannot extract all the information from the images. It is also possible that the CNN is also finding information that is lost in the compression to morphological parameters. Thus, we join the CNN and FCN together, as shown in Fig. \ref{fig:FULL} and explained in detail in \citet{2022A&A...661A..52P}, and pass the concatenated CNN-FCN layer into a discriminator (TN). The CNN and FCN were trained before concatenation and the TN was trained without continuing to train the CNN and FCN. This combined network resulted in a loss of 0.260 and accuracy of 91.7\% at validation, an improvement over the individual CNN and FCN.

As can be seen from these result, the CNN was unable to extract all the information from the images. The FCN lost information in the compression of the images to morphological parameters, or could not fully exploit the morphologies, or both. The combined CNN-FCN allowed at least some of this lost information to be recovered. At test, the CNN-FCN-TN network achieved an accuracy of 88.4\%.

\section{Merger misclassification} 

As well as the network trained at $z < 0.15$, we also trained a network for galaxies at $0.15 \leq z < 0.30$, using the $z < 0.15$ galaxies, edited to appear as higher redshift galaxies. The statistics for these networks is shown in Table \ref{tab:performance}.

\begin{table}
	\caption[]{Performance statistics from the neural networks derived using a class balanced data set (upper) and galaxy classification (lower) \citep{2022A&A...661A..52P}.}
	\label{tab:performance}
	\centering
	\begin{tabular}{ccc}
	\hline
	\hline
		& $z < 0.15$ & $0.15 \leq z < 0.30$\\
	\hline
	Accuracy & 0.884 & 0.850\\
	Recall & 0.863 & 0.890\\
	Precision & 0.901 & 0.899\\
	Specificity & 0.905 & 0.911\\
	NPV & 0.869 & 0.812\\
	\hline
	Total Galaxies & 6965 & 27\,299\\
	Non-merger & 5488 & 18\,518\\
	Merger-candidate & 1477 & 8718\\
	Confirmed merger & 251 & 1858\\
	\hline
	\end{tabular}
\end{table}

We apply these networks to 34\,264 galaxies in NEP and present the classifications in Table \ref{tab:performance} (lower).  10\,195 (29.8\%) of the galaxies were identified as mergers (merger candidates). These were then visually checked, due to the known imperfect nature of the networks. Of the 10\,195 merger candidates, 2109 (20.7\%) were identified as true mergers. This misclassification is worse at $0.15 \leq z < 0.30$ (1858 of 8718 or 21.3\% were true mergers) than at $z < 0.15$ (251 of 1447 or 17.0\% were true mergers). As the negative predictive value (NPV) and recall are worse in the higher redshift network, it was expected the higher redshift merger candidate sample would be more contaminated.

\begin{figure}[t!]
	\resizebox{\hsize}{!}{\includegraphics[]{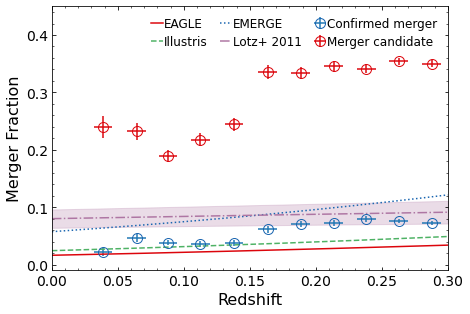}}
	\caption{\footnotesize Merger fraction of the merger candidates (red circles) and visually confirmed mergers (blue circles) as a function of redshift. Merger fractions from \citet[][dot-dashed purple line]{2011ApJ...742..103L}, EAGLE simulation \citep[][solid red line]{2017MNRAS.464.1659Q}, Illustris simulation \citep[][dashed green line]{2015MNRAS.449...49R}, and the pair fraction trend from the EMERGE simulation \citep[][dotted blue line]{2021MNRAS.501.3215O} are also shown.}
	\label{fig:fractions}
\end{figure}

As a simple example of the impact of misclassification, we consider the redshift evolution of the merger fraction in Fig. \ref{fig:fractions}. The merger-candidate merger fraction shows a rapid evolution with redshift which is much higher than the simulation work from EAGLE \citep{2017MNRAS.464.1659Q}, Illustris \citep{2015MNRAS.449...49R} and EMERGE \citep{2021MNRAS.501.3215O}. It is also much higher than the \citet{2011ApJ...742..103L} using mergers selected with traditional methods. The merger fraction for the visually confirmed mergers is inline with other works.

As shown, well trained networks can still produce highly contaminated samples of uncommon or rare astronomical objects. If the classification and contamination rates for galaxy mergers was also found for a sample of $\approx 10^{9}$ galaxies, roughly the expected number of galaxies from Euclid or Rubin-LSST, then $\approx 300\times 10^{6}$ galaxies would be identified as mergers of which only $\approx 60\times 10^{6}$ would be true mergers. It is impractical to check the $\approx 300\times 10^{6}$ merger candidates so care must be taken in the future large scale surveys with how we use the rare object classifications.

\section{Conclusions}
In this work, we have shown that using both images and morphologies in neural networks improves merger classification. We also demonstrated that networks good at identifying rare classes potentially create highly contaminated samples, which has implications for upcoming large area surveys.

\begin{acknowledgements}
We would like to thank the referee for their thoughtful comments that helped improve the quality and clarity of this proceedings.
	
W.J.P. has been supported by the Polish National Science Center project UMO-2020/37/B/ST9/00466 and by the Foundation for Polish Science (FNP).
\end{acknowledgements}

\bibliographystyle{aa}
\bibliography{eas_S11_Pearson}

\end{document}